\documentclass[10pt,nonacm, sigconf]{acmart}
\usepackage[compact]{titlesec} 
\titlespacing{\section}{1pt}{1pt}{1pt} 
\usepackage{booktabs}
\usepackage{multirow}

\usepackage{xspace}
\usepackage{graphicx}
\usepackage{epstopdf}
\usepackage{url}
\usepackage{color}
\definecolor{Orange}{rgb}{0.9,0.5,0}
\definecolor{NavyBlue}{rgb}{0.1, 0.4, 0.8}
\definecolor{Magenta}{rgb}{0.8, 0.1, 0.6}
\definecolor{Green}{rgb}{0.1, 0.8, 0.3}
\definecolor{DarkGreen}{rgb}{0.0, 0.7, 0.2}
\definecolor{Brown}{rgb}{0.4, 0.3, 0.1}
\definecolor{Burgundy}{rgb}{0.5, 0.0, 0.13}
\definecolor{BrightCerulean}{rgb}{0.11, 0.67, 0.84}
\definecolor{BlueViolet}{rgb}{0.33,0.1,0.5}

\AtBeginDocument{%
  \providecommand\BibTeX{{%
    \normalfont B\kern-0.5em{\scshape i\kern-0.25em b}\kern-0.8em\TeX}}}
\copyrightyear{2019}
\acmYear{2019}
\setcopyright{rightsretained}
\acmConference{SIGGRAPH '19 Talks}{July 28 - August 01, 2019}{Los Angeles, CA, USA}
\acmDOI{10.1145/3306307.3328180}
\acmISBN{978-1-4503-6317-4/19/07}
\acmBooktitle{SIGGRAPH '19 Talks, 
  July 28 - August 01, 2019, Los Angeles, CA}
\begin{document}
\title{Emotion Filtering at the Edge}
\author{Ranya Aloufi, Hamed Haddadi, David Boyle}
\affiliation{%
  \institution{Systems and Algorithms Laboratory}
  \institution{Imperial College London}}

\renewcommand{\shortauthors}{}

\begin{abstract}
  Voice controlled devices and services have become very popular in the consumer IoT. Cloud-based speech analysis services extract information from voice inputs using speech recognition techniques. Services providers can thus build very accurate profiles of users' demographic categories, personal preferences, emotional states, etc., and may therefore significantly compromise their privacy. To address this problem, we have developed a privacy-preserving intermediate layer between users and cloud services to sanitize voice input directly at edge devices. We use CycleGAN-based speech conversion to remove sensitive information from raw voice input signals before regenerating neutralized signals for forwarding. We implement and evaluate our emotion filtering approach using a relatively cheap Raspberry Pi 4, and show that performance accuracy is not compromised at the edge. In fact, signals generated at the edge differ only slightly ($\sim$0.16\%) from cloud-based approaches for speech recognition. Experimental evaluation of generated signals show that identification of the emotional state of a speaker can be reduced by $\sim$91\%.
\end{abstract}
\ccsdesc[500]{Embedded systems}
\ccsdesc[300]{Voice-enabled}
\ccsdesc{Security and Privacy}
\ccsdesc[100]{Performance and Utility}

\keywords{Speech Analysis, Voice Synthesis, Voice Privacy, Internet of Things (IoT)}

\maketitle
\vspace{-3mm}
\section{Introduction}
\begin{figure}[t!]
  \centering
  \includegraphics[width=\linewidth]{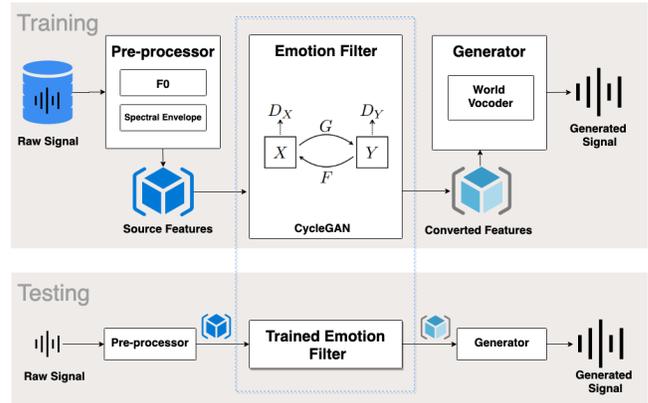}
  \caption{The data-flow of the proposed framework during the training and testing phases}
  \Description{The 1907 Franklin Model D roadster.}
    \label{fig:SysFramework}
\end{figure}
Voice-controlled IoT devices and smart home assistants have gained huge popularity. Seamless interaction between users and services are enabled by speech recognition. Many IoT devices such as home assistants, smartphones, and smart watches have built-in microphones that listen for user commands, and due to resource limitations on edge devices, speech analysis is usually outsourced to cloud services. However, services providers aim to expand their abilities to understand the additional information about the speakers by developing models that process their voice input and detect their current conditions such as emotion classification and analysis of physical and mental wellbeing. They can collect sensitive behaviour patterns from voice input which include various embedded metadata such as "the who, when, where, what and how" that may violate user privacy in numerous ways. They may infer a user's mental state, stress level, smoking habits, overall health conditions, indication of Parkinson's disease, sleep patterns, and levels of exercise~\cite{2014_IoT_Privacy_Peppet}. For instance, Amazon has patented technology that can analyze users' voice to identify emotions and/or mental health conditions. It allows understanding speaker commands and their emotion characterization to provide highly targeted content~\cite{AmazonPatents_2018}. Similarly, Affectiva has developed a multimodal artificial emotional intelligence by combining the  analysis of both face and speech as complementary signals to understand the human  emotion expression~\cite{Affectiva_2019}. Therefore, privacy-preserving speech analysis is playing an especially important role when it comes to advertising content related to physical or emotional states. 

Emotions are a universal aspect of human speech that convey their behaviour. As a consequence of listening to users' voices and monitoring emotions, resulting critical decision-making may affect their life, ranging from fitness trackers for well-being to suitability for recruitment and open many new privacy issues. In this paper, we propose a privacy-preserving architecture for speech analysis and evaluate it to mitigate the privacy risks of the cloud-based voice analysis services. It serves as a mask of the sensitive emotional patterns in the voice input to prevent services providers from monitoring users' emotions that associated with their voice. 

Our proposed solution is a feature learning and data reconstruction framework to bridge the communication between users edge devices and a service provider cloud. It performs emotion filtering in low cost edge devices while still maintaining the usability of the voice input for the cloud-based services. It includes three components: pre-processor, emotion filter, and generator. Firstly, pre-processor extracts the sensitive features to be hidden and then used as a target to train the transformation model. Emotion filter is an embedded-specific model that uses CycleGAN architecture~\cite{3} to transform the voice input style. Finally, the generator uses the output features to re-generate the voice files based on the state-of-the-art vocoder: WORLD~\cite{7}. To evaluate the trade-off between data utility and privacy, the proposed method is tested on an emotion recognition task using the RAVDESS dataset~\cite{11}. The results show that the proposed solution can decrease the accuracy of the emotion recognition application, while affecting the accuracy of speech recognition and speaker identification techniques only minimally. Contributions of this paper can be summarized as follows:
\begin{itemize}
  \item Privacy-preserving emotion filter using CycleGAN that learns how to replace sensitive emotional features of the voice input with corresponding neutral one. 
  \item Implementation and evaluation at the edge versus cloud to show how it can retain the similar performance accuracy in protecting sensitive information at edge and cloud-based approaches.
\end{itemize}

Filtering the affect content of the voice signal is critical task to ensure appropriate protections for users of cloud-based voice analysis services. The proposed framework is the first privacy-preserving emotion filter for voice inputs at edge devices to protect the private paralinguistic information of the speaker. It enables users to protect their sensitive emotional data, while benefiting from sharing their non-sensitive data with cloud-based voice analysis services. In addition, we make our code and results available online.\footnote{\href{https://github.com/RanyaJumah/Embedded-PP-Speech-Analysis}{https://github.com/RanyaJumah/Embedded-PP-Speech-Analysis}}

\section{Related Work}
\textbf{Voice versus Privacy} Voice is considered to be one of the unique biometric information that has been widely used in various IoT applications. It is a rich resource that discloses several possible states of a speaker, such as emotional state~\cite{24}, confidence and stress levels, physical condition~\cite{23,24,27}, age ~\cite{26}, gender, and personal traits. For example, Mairesse et al.~\cite{28} proposed classification, regression and ranking models to learn the Big Five personality traits of a speaker. Previous studies on the voice input privacy have been focused on two main aspects which are the voice-enabled systems breakthroughs and revealing the user's privacy by analyzing their communications. By spoofing voice-based authentication systems, the attackers will have unauthorized access to the private information of these systems users~\cite{13}. Alepis and Patsakis in~\cite{1} presented and analyzed the potential risks of voice assistants in mobile devices, showing how urgent it is to develop privacy-preserving architectures for speech analysis by extracting the distinguishable features from the speech without compromising individual privacy.

\textbf{Edge Computing and Privacy-preserving Deep Learning} One of the primary roles of the edge computing is to filter the data locally prior send it to the cloud which may be used to protect users privacy. In~\cite{osia-2017} a hybrid framework for privacy-preserving analytics is presented by splitting a deep neural network into a feature extractor module on user side, and a classifier module on cloud side. It protects the user privacy by removing the undesired sensitive information from the extracted features results. Generative adversarial networks (GANs)~\cite{4} are one of the deep learning models that have been recently applied to filter the sensitive information from the raw data and regenerate the filtered data. For example, on-device transformation of sensor data was proposed by Malekzadeh et al. in~\cite{Malekzadeh-2019}. They use convolutional auto-encoders (CAE) as a sensor data anonymizer to remove user identifiable features locally and then share the filtered sensor data with specific applications such as daily activities monitoring apps. 

\textbf{Privacy-preserving Voice Analysis on the Edge}
Voice conversion is one of privacy preservation approaches that has been used in the context of speaker identity. For example, VoiceMask is proposed to mitigates the security and privacy risks of the voice input on mobile devices by concealing voiceprints and adding differential privacy~\cite{15}. It sanitized the audio signal received from the microphone by hiding the speaker's identity and then sending the perturbed speech to the voice input apps or the cloud. Nautsch et al. in~\cite{nautsch-2019} investigate the gap in the development of privacy-preserving technologies to protect privacy in the case of speech signals and show the essential need to apply these technologies to protect speaker and speech characterisation in speech recordings.  
\section{System Framework}
\begin{figure}[t!]
  \centering
  \includegraphics[width=\linewidth]{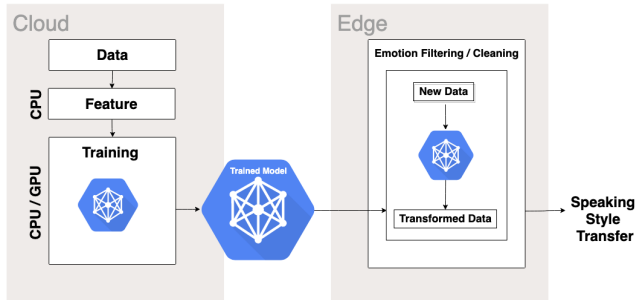}
  \caption{The emotion filter is trained on the cloud, and then the pre-trained filter is used on the edge side for speaking style transformation}
  \Description{The 1907 Franklin Model D roadster.}
    \label{fig:DepFramework}
\end{figure}

Leveraging non-parallel voice conversion (VC) technology~\cite{5}, our framework aims to protect the users' privacy from voice analysis service. The framework consists of three main components: (i) pre-processor, (ii) emotion filter, and (iii) generator. The description of the proposed framework is presented in Figure~\ref{fig:SysFramework}.

\textbf{Pre-processor}
The raw voice input is pre-processed to extract the distinguishing signal representation by performing transformations functions to the voice input, and using the resulting outcomes as labels ~\cite{6}. Prosody features such as spectral envelope (SPs) are the most effective features in emotion recognition tasks~\cite{2016_Trigeorgis} which can be computed directly from the signal by applying specific transformation functions to minimize the computational overhead and extract these specific features. WORLD vocoder ~\cite{7} is used to extract these specific features at frame-level (a frame denotes a number of samples with the same time-stamp, one per channel) from both the source and corresponding target signals. 

\textbf{Emotion Filter}
To learn the sensitive representations in the voice input, CycleGAN-based speaking style conversion is used to transform the raw voice emotional features to corresponding normal one. CycleGAN~\cite{3} is a custom model of GANs that uses two generators and two discriminators. By considering X and Y as different domains that generators task to convert from X to Y and vice versa. Generator (G) maps from domain X to Y, and generator (F) maps from Y to X. In addition, two adversarial discriminators D (X) and D (Y), where D (X) aims to distinguish between objects in X domain and output objects from F (Y), and D (Y) aims to discriminate between (Y) and the output of G (X). It has introduced to overcome the difficulty of preparing paired dataset in style conversion applications ~\cite{3}. Therefore, changing emotion style by using CycleGAN will help to transfer between emotions and neutral speaking style without paired training data.

Consistent with TinyML \cite{TinyML_2019}objectives to trade-off between the machine learning accuracy and resource efficiency and optimize the performance cost of data analysis on constrained platforms such as IoT devices, we propose to implement the emotion filter on the edge to enable on-device privacy-preserving speech analysis. A pre-trained CycleGAN model is frozen by combining the model graph structure with its weight to create an embedded version that can fit on the edge devices for features transformation task.  

\textbf{Generator}
The WORLD synthesis algorithm is used to re-generate a high-quality synthesized voice input by using the output features from the emotion filtering model. The generated voices are able to preserve the content of the voice input and project away sensitive representations such as emotional patterns.

In this way, the sensitive patterns in the voice input will not disclose to cloud-based voice input services providers and they will have only access to the synthesized voice. The output of the proposed framework will protect the speaker privacy by preserving the linguistic content and hiding the private non-linguistics content (emotional patterns).

\section{Experimental Evaluation}

\begin{table}[]
\caption{A comparison of the accuracy between raw and generated voice across speech and speaker recognition tasks.}
\label{tab:Accurecy_Table}
\resizebox{\columnwidth}{!}{%
\begin{tabular}{|l|c|c|}
\hline
\multicolumn{1}{|c|}{\multirow{2}{*}{}} & \textbf{Speech Recognition}       & \textbf{Speaker Recognition}       \\ \cline{2-3} 
\multicolumn{1}{|c|}{}                  & \textbf{Word Error Rate (WER \%)} & \textbf{Equal Error Rate (EER \%)} \\ \hline
Raw Voice                               & 5.27                              & 0.06                               \\ \hline
NVIDIA Quadro P1000                     & 20.36                             & 0.120                               \\ \hline
Intel Core i7                           &   20.66                                &  0.121                                  \\ \hline
ARM Cortex A-72                         & 20.67                             & 0.124                               \\ \hline
\end{tabular}%
}
\end{table}
\begin{figure}[t!]
  \centering
  \includegraphics[width=1\linewidth]{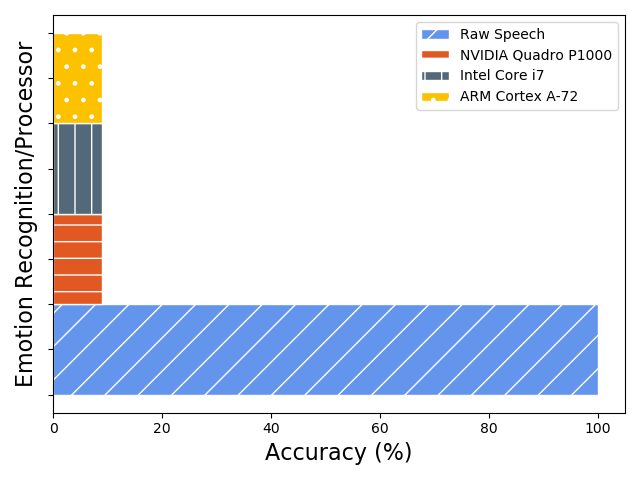}
  \caption{The emotion recognition accuracy of the raw and generated voices: similar performance accuracy by NVIDIA Quadro P1000, Intel Core i7, and ARM Cortex-A72 processor in hiding sensitive emotional patterns.}
  \Description{The 1907 Franklin Model D roadster.}
    \label{fig:emotion_recogntion}
\end{figure}
\begin{figure*}[]
  \centering
  \includegraphics[width=\linewidth]{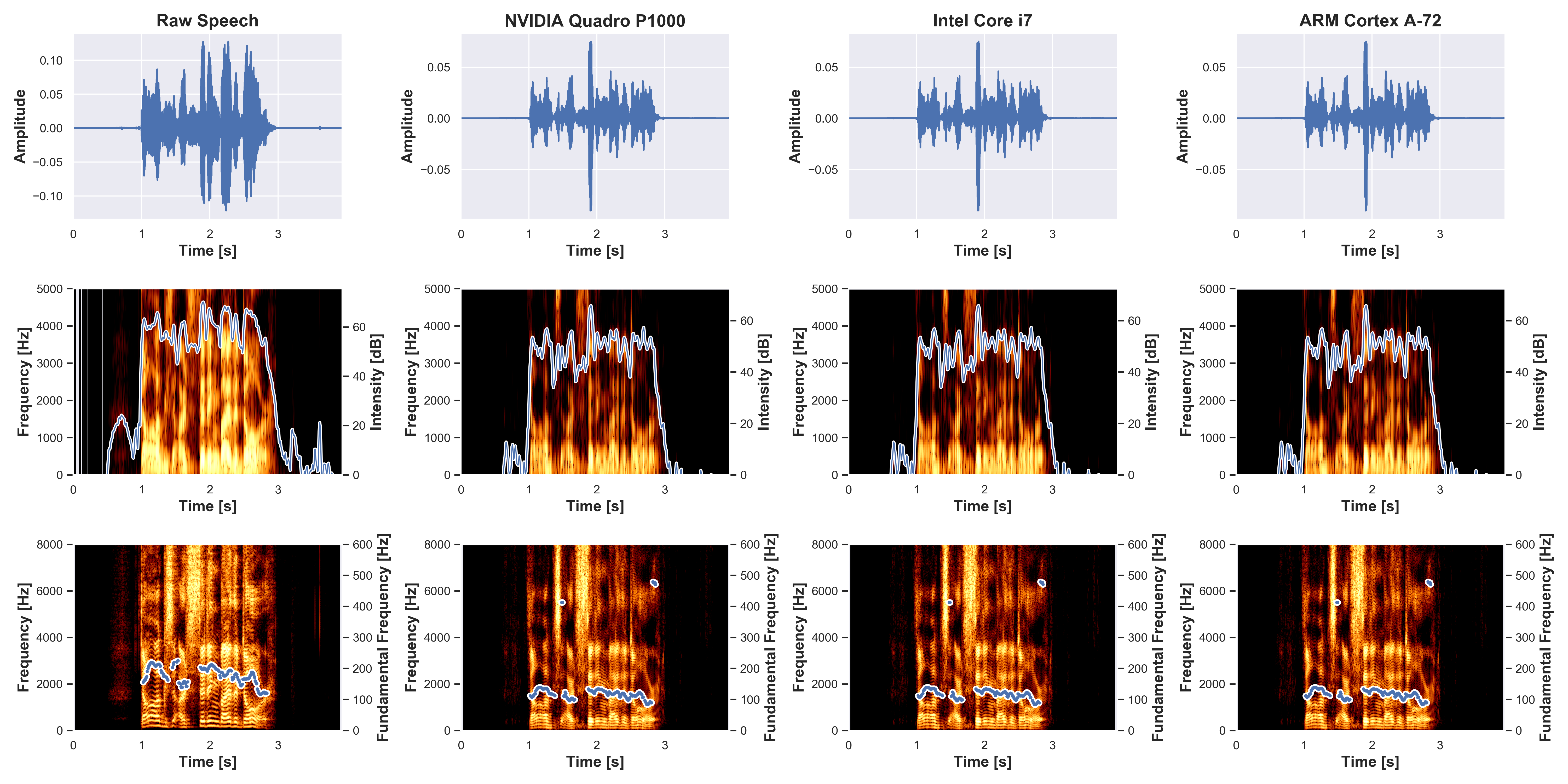}
  \caption{Spectrogram analysis of the raw (happy) and generated waveform (neutral) in term of (top)amplitude (the size of the oscillations of the vocal folds), (middle)intensity (acoustic intensity), and (bottom)fundamental frequency (vocal fold vibration property)}
  \Description{The 1907 Franklin Model D roadster.}
    \label{fig:Spectrogram}
\end{figure*}

The sensitive information to be hidden is the speaker's emotion. As our framework aims to decrease the emotion recognition accuracy while maintaining the speech and speaker recognition accuracy, the following subsections describe the experimental setting, selected speech analysis tasks for evaluation, and evolution results. 

\subsection{Experimental Setting}
We conduct the experiment by running the proposed framework over NVIDIA Quadro P1000 with GP107 graphics processor and 4 GB memory, MacBook Pro with 2.7 GHz Intel Core i7 processor and 8 GB memory and Raspberry Pi4 with ARM Cortex-A72 and 4 GB memory.  The experiment is executed over speech audio-only files in .wav format from the Ryerson Audio-Visual Database of Emotional Speech and Song (RAVDESS)~\cite{11} with 48kHz/16-bit sampling rate. It contains recordings from 24 professional actors (12 female, 12 male), vocalizing two lexically-matched statements in a neutral North American accent with seven speech emotions: 0 = neutral, 1 = calm, 2 = happy, 3 = sad, 4 = angry, 5 = fearful, 6 = disgust, 7 = surprised. A subset of this dataset is used to evaluate the effectiveness of the proposed framework. We select 118 files where 96(training) and 22(testing) files with three emotions: neutral, happy, and angry. It organised as follows (24 * 2) emotion files and (24 * 2) neutral files. To overcome the model over-fitting, different texts have been choosing for training and testing sets.

As depicted in Figure~\ref{fig:DepFramework}, the training phase begins at the cloud end (NVIDIA Quadro P1000 or MacBook Pro) by downsampling the .wav files to 16 kHz to preserve the signal content information. Then, acoustic parameters spectral envelopes (SEs) for each logarithmic fundamental frequency (log F0) are extracted as prosody features which are the most related feature to emotion recognition. These features are mapped from utterances spoken in an emotional style to corresponding features of neutral utterances using CycleGAN with similar network architecture in ~\cite{5}. The emotion filter is trained under 7500 iterations with learning rates of 0.0002 for the generator and 0.0001 for the discriminator. On the edge side (Raspberry Pi 4), the pre-trained emotion filter is exported to apply on-device emotion filtering. Then, the raw voice signal is pre-processed to extract the prosody features. The emotionless speaking style is achieved by using the pretrained filter to convert these features in the raw signal. Finally, the outputs of the conversion phase are converted to neutral speech waveforms by using the WORLD synthesizer.

\subsection{Speech Analysis Tasks}
We conducted objective evaluations of the generated voice for three speech analysis tasks. Firstly, the re-generated .wav file is evaluated by the trained emotion recognition model on RAVDESS dataset to identify the emotion state of the sanitized voices. Then, we perform speech and speaker recognition on the sanitized voices and evaluate the accuracy. The tools used in the three tasks evaluation is described as follows.

\textbf{Speech Recognition} The IBM Watson speech-to-text service with speech recognition capabilities is used to convert the generated speech into text ~\cite{20}. The performance of the speech recognition on the sanitized voices is measured by the word error rate (WER), which is a common metric of the speech recognition performance to measures the difference in the word level between two spoken sequences.

\textbf{Speaker Recognition} To ensure that the proposed system is highly confident that person is speaking and has been correctly identified, a VGG speaker recognition model on VoxCeleb2~\cite{18} has been used. All audio files are converted to 16-bit streams at a 16 kHz for consistency. The accuracy of speaker recognition is measured by the equal error rate (EER), which is the rate at which both acceptance and rejection errors are equal.

\textbf{Emotion Recognition} To automatically identify the emotional state of the users, an emotion classification model based on RAVDESS dataset has been used to predict 7 emotion classes which are the following: 0 = neutral, 1 = calm, 2 = happy, 3 = sad, 4 = angry, 5 = fearful, 6 = disgust, 7 = surprised ~\cite{19}. The accuracy of emotion recognition is defined as the success rate of correctly identified emotions.

\subsection{Evaluation and Discussion}
From the experiments results, we can summaries the following:

\textbf{Results accuracy and privacy}
We compare the accuracy results from the speech and speaker recognition models on raw and transformed voices, and it shows that the utility of the signal retains accepted, while the emotion recognition accuracy is sharply decreased, see Figure \ref{fig:emotion_recogntion}. The evaluation of the speech recognition performance is done using the average of WER which is 20.36 \%, 20.66 \%, and 20.67 \% in NVIDIA Quadro P1000, Intel Core i7, and ARM Cortex A-72 respectively. In addition, the speaker recognition performance is measured by EER and the average of the error rate is $\sim$0.12 in all three platforms, as shown in Table \ref{tab:Accurecy_Table}. As a result, the proposed framework has insignificant difference in the performance accuracy across edge and cloud-based resources. However, the speech recognition accuracy will be improved by increasing the dataset, refining the features set, and manipulate the model architecture.

\textbf{Model optimization on the edge}
With relatively cheap ARM Cortex-A72 board device, we show that the proposed framework can be implemented with similar performance accuracy as on NVIDIA Quadro P1000 and Intel Core i7. Spectrogram analysis of the raw and transformed speech is illustrated in Figure \ref{fig:Spectrogram}, demonstrates that there are similar changes on the amplitude, intensity, and fundamental frequency of the transformed speech using cloud and edge resources which lead to alike accuracy performance over speech analysis tasks. However, optimizing the model will be considered by implementing different approaches such as weight pruning, compression, and quantization to enhance the model performance on the edge.

\begin{figure}[t!]
  \centering
  \includegraphics[width=\linewidth]{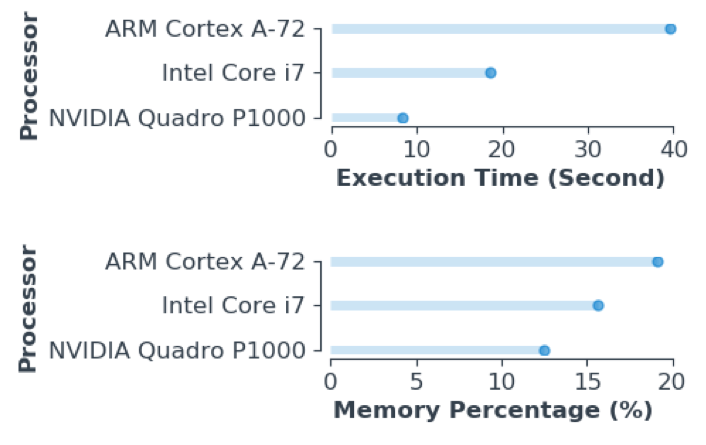}
  \caption{A comparison between the execution time and memory usage of running the model in NVIDIA Quadro P1000, Intel Core i7, and ARM Cortex-A72 }
  \Description{The 1907 Franklin Model D roadster.}
    \label{fig:System_utility}
\end{figure}

\textbf{Resource limitation and scalability}
Computational performance is limited by various resources constraint such as memory capacity. Figure \ref{fig:System_utility} compares the execution time and memory usage of the emotion conversion model running on the NVIDIA Quadro P1000 and Intel Core i7 versus the ARM Cortex-A72. However, NVIDIA Quadro P1000 and Intel Core i7 outperforms ARM Cortex-A72 in the execution time. Precisely, ARM Cortex-A72 implementation consumes twice as much time as that of the Intel Core i7. We managed to significantly reduce the execution time and memory usage of running the proposed framework on edge devices.

\begin{figure}[t!]
  \centering
  \includegraphics[width=1\linewidth]{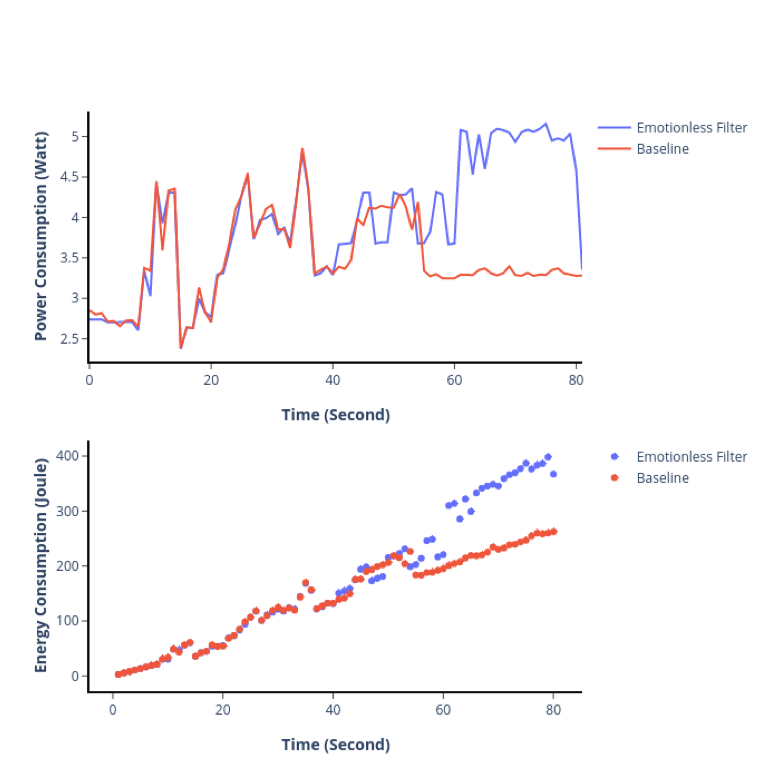}
  \caption{Power and Energy Consumption}
  \Description{The 1907 Franklin Model D roadster.}
    \label{fig:privacy-cost}
\end{figure}

\textbf{Privacy overhead on the edge}
We performed a privacy overhead analysis for the emotion filter on the edge devices. We use two type of speech analysis experiments: one integrated with the emotion filter and a baseline experiment without emotion filter, as described in Table \ref{tab:Privacy-overhead}. In the first experiment, we disable the emotion filter and measure the overhead purely incurred by configuring the edge device (raspberry pi), loading the .wav file, and uploading the file to the cloud. The second experiment type measures the overhead purely incurred by running the speech analysis with emotion filter. Precisely, we measure the overhead by: (1) configuring the edge device, (2) loading .wav file, (3) pre-processing, filtering, and generating the .wav file, and (4) uploading the file to the cloud.  Figure \ref{fig:privacy-cost} shows the experiment results. Raspberry pi needs about 40 second for booting up. The ${\Delta}$ average power consumption is 0.45 Watt and ${\Delta}$ average energy consumption is 31.2 Joule. The baseline latency is about 20 second, while the emotion filter latency is 41 second.

\section{Discussion and Future Work}

In this paper we presented a framework for privacy preserving speech analytics which consists of a pre-processor, an emotion filter, and a generator. It will protect the user privacy by hiding the undesired sensitive information from the extracted features by transforming the features correspond to emotional pattern while retaining the features correspond to speech content and speaker identity unchanged. Therefore, on the cloud side, only non-sensitive filtered features can be inferred such as linguistics content. Evaluating our framework by distribution the training and testing execution between the edge and the cloud, we achieved high decrease in the emotion recognition task accuracy by $\sim$91\%, while slightly decreasing for other tasks such as speech and speaker recognition. 
Protection the users' privacy in speech analysis is a very challenging task. The challenge is how to sanitize the speech without decreasing the speech recognition accuracy. We will focus on extending the proposed framework by including speech content filter to prevent similar outcomes using other techniques, such as sentiment analysis to strengthen the user privacy. In addition, we will include speech analysis in-the-wild emotional dataset and further investigations of for privacy-preserving deep learning architecture. 

\begin{table}[]
\caption{Privacy Overhead Analysis Experiments}
\label{tab:Privacy-overhead}
\resizebox{\columnwidth}{!}{%
\begin{tabular}{|l|l|l|}
\hline
Time & \multicolumn{1}{c|}{\textbf{Baseline}} & \multicolumn{1}{c|}{\textbf{Emotionless Filter}}             \\ \hline
T0   &  Pi on                               &  Pi on                                                 \\ \hline
T1   &  Load .wav                           &  Load .wav                                             \\ \hline
T2   &  Cloud uploading                     &  Preprocessor (PP), emotion filter (EF), generator (G) \\ \hline
T3   &                                        &  Cloud uploading                                       \\ \hline
\end{tabular}%
}
\end{table}

\bibliographystyle{ACM-Reference-Format}
\balance
\bibliography{sample-base}

\end{document}